\documentclass{ws-ijmpa_Standard}
\usepackage{graphicx}
\begin{document}

\markboth{I. Caprini, G. Colangelo, H. Leutwyler}{Pion-pion interaction}

\catchline{}{}{}{}{}

\title{\bf THEORETICAL ASPECTS OF THE PION-PION INTERACTION }

\author{I. CAPRINI\footnote{Speaker at the International Conference on QCD
    and Hadronic Physics, June 16-20 2005, Beijing.}} 
\address{National Institute of Physics and Nuclear Engineering,
 Bucharest,  R-077125 Romania}

\author{G. COLANGELO,  H.  LEUTWYLER }
\address{Institute for Theoretical Physics, University of 
Bern,
Sidlerstr. 5, CH-3012 Bern, Switzerland}  


\maketitle


\begin{abstract}
  We give a brief review of the theoretical description of low energy
  pion-pion scattering by the combined use of Chiral Perturbation Theory
  and Roy equations, an update of the Regge parametrization of $\pi\pi$
  cross sections at high energies, and a short discussion of the scalar
  radius of the pion.
\end{abstract}


\section{Low-energy scattering: ChPT and Roy equations}
The structure of the $\pi\pi$ interaction at low energies is strongly
constrained by the fact that the pion is the Goldstone boson of a
spontaneously broken approximate symmetry of QCD.\cite{Weinberg} The
scattering amplitude can be calculated in a systematic way in the framework
of Chiral Perturbation Theory (ChPT),\cite{GL} by an expansion which
converges very rapidly near the center of the Mandelstam triangle. However,
the convergence becomes slow as one approaches the unitarity cuts, and
already at threshold the direct application of chiral expansions is not 
satisfactory.

The dispersion relations determine the structure of the amplitude in terms
of physical region absorptive parts and two subtraction constants, which
can be identified with the S-wave scattering lengths, $a_0^0$ and $a_0^2$.
By projecting the fixed-$t$ dispersion relations onto partial waves and
using unitarity, one obtains a set of integral equations for the phase
shifts, the Roy equations.\cite{Roy}

In the early applications of Roy equations the subtraction constants were
taken from experiment, and had large uncertainties. An important step
forward was to determine them theoretically, by combining ChPT inside the
convergence region with dispersion relations, which led to a very precise
theoretical prediction:\cite{CGL}
\begin{equation}
a_0^0=0.22\pm 0.005\,, \quad\quad a_0^2=-0.0444\pm 0.0010 \,.
\end{equation}
This range is represented by the small red region in Fig. 1, where the
black lines define the universal band\cite{ACGL} imposed by Roy equations,
the points are the ChPT calculations in the tree, one-loop, and two-loop
approximation (see Ref.4 for references), and the blue dashed lines are
obtained by a Roy analysis\cite{DGFGS} that does not make use of chiral
perturbation theory. We also show the recent, unquenched lattice result for
$a^2_0$ obtained by NPLQCD,\cite{NPLQCD} as well as the range for $a^0_0$
and $a^2_0$ that corresponds to the values for the LECs given in Ref.8. The
experimental results are from BNL E865,\cite{Pislak} CERN
DIRAC,\cite{DIRAC} and CERN NA48/2\cite{NA48} experiments.
\begin{figure}  \vspace*{8pt}
\centerline{\psfig{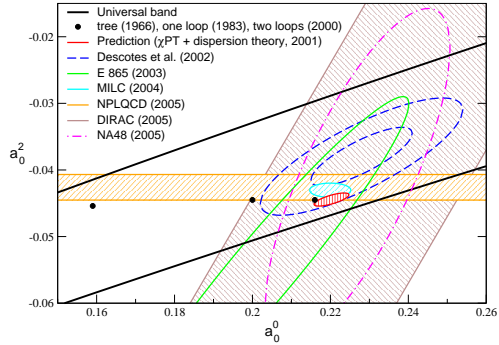}}
\caption{ Theoretical and experimental status of the S-wave scattering
  lengths (details in the text). } 
\end{figure}

Having precise values for the subtraction constants, the threshold
parameters, the S and P-wave phase shifts below 0.8 GeV, and the coupling
constants of the effective chiral SU(2)$\times$SU(2) Lagrangian relevant
for $\pi\pi$ scattering could be determined to high
accuracy.\cite{CGL}$^,$\cite{ACGL} Below 0.8 GeV the influence of the high
energies in the Roy equations is very small. This was shown in Ref.13,
where we explored the sensitivity of the Roy solutions with respect to the
high energy input. We solved Roy equations using at high energy, instead of
the Regge model adopted in Ref. 5, an alternative Regge parametrization
proposed recently.\cite{PY} Our analysis\cite{CCGL} shows that the
predictions for the S-wave scattering lengths and the isoscalar S-wave and
P-wave phase shifts are practically unchanged, while the exotic S-wave
phase shift is modified by only 1.4$^\circ$ at 0.8 GeV. For a detailed
discussion see Ref. 13.
\section{High-energy scattering: Regge parametrization} 
The Regge parametrization of $\pi\pi$ amplitudes was discussed in the
seventies, but afterwards the interest in this subject diminished.
Recently, a Regge analysis of $\pi\pi$ amplitudes was presented in Ref. 12.
We are currently performing a new analysis,\cite{CCL} exploiting
factorization of Regge residua and dispersive sum rules. The motivation of
the study is to have an improved input for solving Roy equations above 0.8
GeV, in the validity region $\sqrt{s}\leq 1.15\,\mbox{GeV}$.\cite{Roy}
Below we give only a few results on the Regge parametrization of the total
cross sections.

Using the notations adopted in Ref. 15 for $\pi^\pm N$, $NN$ and $\bar NN$
scattering, we express the $\pi\pi$ total cross sections as
\begin{equation}\label{sigmatot} \sigma_{\pi^+\pi^\pm} =
  B\,\ln^2(s/s_0) +Z_{\pi\pi}+Y_{1\pi\pi}\,
  (s_1/s)^{\eta_1}\mp Y_{2\pi\pi}\,(s_1/s)^{\eta_2}\,.\end{equation}
The first two terms are the contribution of the Pomeron, the last two the
contribution of the $f$ and $\rho$ Regge poles, respectively.  $Z$ and $Y$
denote the Regge residua supposed to satisfy factorization, and  
$\eta_j$ are related to the trajectories intercepts.\cite{PDG2004}

In Figs. 2 we present our results for the total cross sections with
definite isospin in the t-channel: $I_t=0$, which receives contribution
from the Pomeron and $f$, and $I_t=1$, dominated by the $\rho$ Regge pole.
The sum of the low partial waves at lower energies is indicated for
comparison. The red bands are obtained by using the fits\cite{PDG2004} of
$\pi N$ and $N N$ data above 5 GeV and the factorization of Regge residua.
The blue and green bands represent the parametrizations considered in Ref.
5 and Ref. 12, respectively. The bands denoted as ``our estimates'' are
obtained, for $I_t=0$ channel, by ascribing an extrapolation error to the
parameters derived from factorization above 5 GeV and, for $I_t=1$ channel,
by applying Olsson sum rule\cite{CCGL} (in this case the results given by
factorization have large uncertainties). Our results show that in Ref.5 the
$I_t=0$ contribution was slightly underestimated, while in Ref. 12 the
authors take too small a value for the $\rho$ residue. Details will be
reported elsewhere.\cite{CCL}
 \begin{figure}
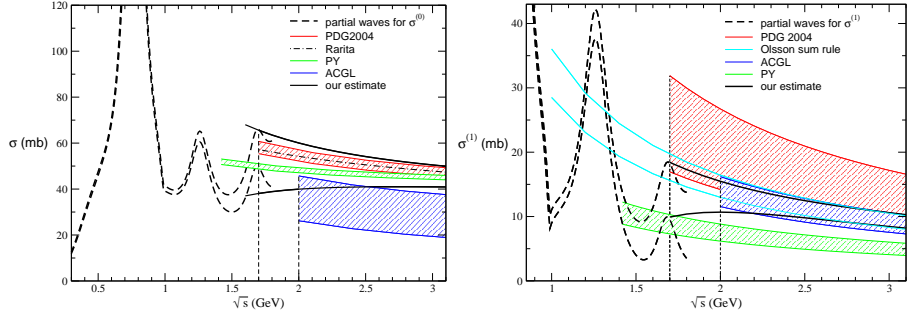
 \vspace*{3pt}
 \includegraphics[width=5.85cm]{sigma0tCCLnew.eps}
 \includegraphics[width=6cm]{sigma1tCCLcolor.eps}
\vspace*{5pt}
\caption{ Total cross section in the channels with  $I_t=0$ (left) and
  $I_t=1$ (right). }  
\end{figure}
\section{Scalar radius of the pion}
The scalar radius of the pion is an important quantity in ChPT, because it
is related to an effective coupling constant, $\bar{\ell}_4$, that
determines the first nonleading contribution in the chiral expansion of the
pion decay constant.  A first crude estimate for the scalar radius in ChPT
reads\cite{GLform}: $\langle r^2\rangle_s^\pi = 0.55\pm 0.15
\,\mbox{fm}^2$.  An improved result, $\langle r^2\rangle_s^\pi = 0.61\pm
0.04 \,\mbox{fm}^2$, was obtained from dispersion theory and two-channel
unitarity for the scalar form factor $ \Gamma_\pi(s)$ (see Ref. 17 for a
recent discussion and references to earlier works).  This result was
questioned in Ref. 18, where the author used a single channel Omn\`es
representation of the scalar radius in terms of the phase
$\delta_\Gamma(s)$ of the form factor, to advocate a larger value: $\langle
r^2 \rangle_s^\pi=0.75\pm0.07\,\mbox{fm}^2$.  However, as we emphasized in
Ref. 17, the estimate\cite{Yndurain1} of the phase $\delta_\Gamma(s)$
ignores an ambiguity of $\pm \pi$ in the Watson theorem, which can be the
resolved only by the explicit inclusion of inelastic channels in the
Mushkhelishvili-Omn\`es formalism.
 
In Ref.19 the author invokes perturbative QCD in favor of a large phase
$\delta_\Gamma (s)$. To leading order in $\alpha_s$, neglecting quark and
pion masses in the propagators, one obtains\footnote{Useful discussions
  with D. Pirjol on this subject are aknowledged.}  for large spacelike
momenta $Q^2>0$:
\begin{equation}\label{QCD}
  \Gamma_\pi(Q^2) \sim   \frac{4 \pi f_\pi^2 \alpha_s(Q^2)}{Q^2}
  \int\limits_0^1 {\rm d}\xi \int\limits_0^1 {\rm d} \eta
  \left[\bar{m}_u^2(Q^2)\,\frac{ \phi(\xi) \, \phi(\eta)}{ \xi (1-\eta)^2}+
    M_\pi^2 \, \frac{ \phi(\xi) \, \phi_p(\eta) }{ \xi (1-\eta)}\right],
\end{equation}
where $ \phi (\xi)=6 \xi (1-\xi)$ and $\phi_p(\xi)=1$ are the twist-2 and
twist-3 light-cone distribution amplitudes, respectively.\cite{Beneke} Both
terms in Eq. (\ref{QCD}) contain an end-point logarithmic divergence, which
is usually replaced by $\ln (Q^2/\Lambda_{\rm QCD}^2)$.\cite{Beneke} In
Ref. 19 the author keeps only the first term in (\ref{QCD}) and claims that
around 1 GeV the phase $\delta_\Gamma(s)$ is much larger than its
asymptotic limit $\pi$. But the first term in (\ref{QCD}) vanishes faster
than the second one for $m_u\to 0$ and represents, in comparison, a small
correction. Therefore, the arguments put forward in Ref. 19 are based on an
incomplete calculation.  Moreover, the logarithmic singularity makes the
specific predictions from QCD doubtful in this case. A complete discussion
will be given elsewhere.

\section*{Acknowledgements}
We thank B. Ananthanarayan and J. Gasser for a nice and fruitful
collaboration. I. Caprini is grateful to the conference organizers for the
invitation.

\end{document}